\documentclass{emulateapj}

\usepackage{float}
\usepackage{amsmath}
\usepackage{epsfig,floatflt}
\usepackage{subfigure}

\newcommand{\BA}{\mathbf{A}}

\newcommand{\BC}{\mathbf{C}}
\newcommand{\Ba}{\mathbf{a}}

\newcommand{\Bs}{\mathbf{s}}
\newcommand{\BS}{\mathbf{S}}
\newcommand{\Bn}{\mathbf{n}}
\newcommand{\BN}{\mathbf{N}}

\newcommand{\Bd}{\mathbf{d}}

\newcommand{\id}{\mathbf{1}}
\newcommand{\Bx}{\mathbf{x}}
\newcommand{\By}{\mathbf{y}}

\newcommand{\Bu}{\mathbf{u}}
\newcommand{\Bv}{\mathbf{v}}

\newcommand{\Cell}{C_{\ell}}

    \newcommand{\qed}{\nobreak \ifvmode \relax \else
          \ifdim\lastskip<1.5em \hskip-\lastskip
          \hskip1.5em plus0em minus0.5em \fi \nobreak
          \vrule height0.75em width0.5em depth0.25em\fi}

\begin{document}

\title{A Markov Chain Monte Carlo Algorithm for analysis of low
  signal-to-noise CMB data} 

\author{J. B.  Jewell\altaffilmark{1, 3, 4}, H.\ K.\
  Eriksen\altaffilmark{2,5,6}, B.\ D.\ Wandelt\altaffilmark{7,8}, 
 I.\ J.\ O'Dwyer\altaffilmark{3},  Greg Huey\altaffilmark{3}, and K. M.
  G\'{o}rski\altaffilmark{3,4,9}}

\altaffiltext{1}{email: Jeffrey.B.Jewell@jpl.nasa.gov}
\altaffiltext{2}{email: h.k.k.eriksen@astro.uio.no}

\altaffiltext{3}{Jet Propulsion Laboratory, 4800 Oak
  Grove Drive, Pasadena CA 91109} 

\altaffiltext{4}{California Institute of Technology, Pasadena, CA
  91125} 

\altaffiltext{5}{Institute of Theoretical Astrophysics, University of
Oslo, P.O.\ Box 1029 Blindern, N-0315 Oslo, Norway}

\altaffiltext{6}{Centre of
Mathematics for Applications, University of Oslo, P.O.\ Box 1053
Blindern, N-0316 Oslo}

\altaffiltext{7}{Department of Physics, University of Illinois,
  Urbana, IL 61801}

\altaffiltext{8}{Astronomy Department, University of Illinois at
  Urbana-Champaign, IL 61801-3080}

\altaffiltext{9}{Warsaw University Observatory, Aleje Ujazdowskie 4, 00-478 Warszawa,
  Poland}

\date{Received - / Accepted -}

\begin{abstract}
  We present a new Monte Carlo Markov Chain algorithm for CMB analysis
  in the low signal-to-noise regime. This method builds on and
  complements the previously described CMB Gibbs sampler, and
  effectively solves the low signal-to-noise inefficiency problem of
  the direct Gibbs sampler. The new algorithm is a simple
  Metropolis-Hastings sampler with a general proposal rule for the
  power spectrum, $C_{\ell}$, followed by a particular deterministic
  rescaling operation of the sky signal, $\mathbf{s}$. The acceptance
  probability for this joint move depends on the sky map only through
  the difference of $\chi^2$'s between the original and proposed sky
  sample, which is close to unity in the low signal-to-noise regime.
  The algorithm is completed by alternating this move with a standard
  Gibbs move. Together, these two proposals constitute a
  computationally efficient algorithm for mapping out the full joint
  CMB posterior, both in the high and low signal-to-noise regimes.
\end{abstract}


\maketitle

\section{Introduction}

Since the detection of anisotropy in the Cosmic Microwave Background
(CMB; Smoot et al.\ 1992), there has been an emphasis on likelihood or Bayesian methods
for the inference of cosmological parameters and their error bars, or
more generally, their confidence intervals. CMB analysis is most
suitably addressed in a Bayesian, as opposed to frequentist,
framework, simply because the observed microwave sky is interpreted as
a single realization of a spatial random process. 

Early measurements of the CMB were limited to signal to noise ratios
of order unity at relatively low angular scales, where direct
evaluation of the likelihood for the power spectrum or cosmological
parameters is possible. However, the ${\cal O}(N^{3})$ scaling of
computational expense with pixel number $N$ prohibits direct
likelihood evaluation for current and future CMB observations.
Motivated by the scientific potential of CMB data with increasingly
high spatial resolution, yet beset with systematics including partial
sky coverage and foregrounds, an iterative method of sampling from the
Bayes posterior, using a special case of Markov Chain Monte Carlo
(MCMC) known as Gibbs sampling, was introduced by \citep{jewell:2002,jewell:2004}.
The method was later independently discovered and applied to COBE data
\citep{wandelt:2004}, numerically extended to high-resolution on the
sphere \citep{eriksen:2004}, applied to analysis of the WMAP
\citep{bennett:2003, hinshaw:2007, page:2007} data \citep{odwyer:2004,
  eriksen:2007a, eriksen:2007b}, as well as generalized to include
inference of foreground model parameters \citep{eriksen:2008a,
  eriksen:2008b}.

While Gibbs sampling provably converges to the Bayes posterior over
the entire range of angular scales probed by the data, the run-time
required to generate enough independent samples at the low
signal-to-noise, small angular scale regime was found to be
prohibitive \citep{eriksen:2004}. The reason for this is that typical
variations in the power spectrum from one sample to the next are
determined by cosmic variance alone, whereas the posterior itself is
given by both cosmic variance and noise.  This results in a long
correlation length in the sequence of spectra in the low signal to
noise regime, thus requiring a very long run time to generate a
sufficient number of independent samples.  

In this paper we generalize the original Gibbs sampling algorithm to
include a new type of MCMC step alternating with standard Gibbs
sampling, which solves this problem of slow probabilistic convergence
in the low signal to noise regime.  This method therefore makes
possible an exact Bayesian approach to CMB analysis over the entire
range of angular scales probed by current and future experiments.

The paper is organized as follows. We first review the CMB Gibbs
sampler, and describe the associated numerical difficulties in
analysis at small angular scales.  We then introduce the new MCMC step
to the Markov chain, designed specifically to allow large variations
in the high-$\ell$ CMB spectrum, precisely where the signal to noise
is $\le 1$. We derive the required Metropolis-Hastings acceptance
probability correctness in Appendix \ref{app:proof}, and numerically
demonstrate the method in Section \ref{sec:simulations}, for both
temperature and polarization. Finally, we summarize and conclude in
Section \ref{sec:conclusions}.

\section{Review of Gibbs Sampling}

\subsection{The Joint Posterior}

We begin by assuming that the observed data may be modelled by a
signal and a noise term,
\begin{equation}
\Bd = \BA \Bs + \Bn,
\end{equation}
where $\Bd$ is a vector containing the data (at every pointing of the
detectors), the matrix $\BA$ involves both pointing and beam
convolution (and where for this paper we will assume symmetric beams
and neglect the details of this operation), and $\Bn$ is additive
noise (here in the pixel domain).  We assume both the CMB signal and
noise to be Gaussian random fields with vanishing mean and covariance
matrices $\BS$ and $\BN$, respectively. In harmonic space, where $\Bs
= \sum_{\ell, m} a_{\ell m} Y_{\ell m}$, the CMB temperature
covariance matrix is given by $\textrm{C}_{\ell m, \ell' m'} = \langle
a_{\ell m}^* a_{\ell' m'}\rangle = C_{\ell} \delta_{\ell \ell'}
\delta_{m m'}$, $\Cell$ being the angular power spectrum.  A
generalization to polarization merely requires the replacement of the
signal matrix diagonal elements with $3 \times 3$ matrices of the form
\begin{equation}
\BC_{l} = \left[ \begin{array}{ccc}
C_{l}^{TT} & C_{l}^{TE} & C_{l}^{TB} \\
C_{l}^{ET} & C_{l}^{EE} & C_{l}^{EB} \\
C_{l}^{BT} & C_{l}^{BE} & C_{l}^{BB} \end{array} \right]
\end{equation}
For the discussion in this section, we focus on the temperature case,
but note that the generalization to polarization is straightforward
and discussed by \citet{larson:2007}.

Given these asumptions, our goal is to quantify what has been learned
about the underlying power spectrum of the CMB given the data, or how
well the data constrain the cosmological parameters.  One proceeds
then, in a Bayesian framework, by writing down the posterior given the
data,
\begin{equation}
P(C_{\ell} | \Bd) \propto \mathcal{L}(\Bd | C_{\ell} ) P(C_{\ell}).
\end{equation}
Here $\mathcal{L}(\Bd | C_{\ell})$ is the likelihood and $P(C_{\ell})$
is a prior on $C_{\ell}$. 

In order to derive the functional form of the likelihood, one imagines
randomly choosing any relevant model [here a power spectrum drawn from
$P(C_{\ell})$], and asks what sequence of effects needs to be modeled
in order to simulate the data. Here, simulation is understood as
conditioning on the chosen model, and leads to a joint density
\begin{eqnarray}
P(\Bd,\Bs,C_{\ell}) & = & P(\Bd,\Bs | C_{\ell}) P(C_{\ell}) \nonumber \\
& = & P(\Bd | \Bs) P(\Bs | C_{\ell})  P(C_{\ell})
\end{eqnarray}
where the last line follows directly from our data model through the
assumption of additive noise. Specifically, the factors in the above
are
\begin{eqnarray}
-2 \log P(\Bs | C_{\ell} ) & = & \Bs^{t} \BC^{-1} \Bs - \log |\BC| \nonumber \\
-2 \log P(\Bd | \Bs) & = & -(\Bd-\Bs)^{t} \BN^{-1}(\Bd-\Bs) - \log |\BN|
\end{eqnarray}
which follow from the assumption that both the signal and noise are
independent Gaussian processes.

The idea of a ``simulation chain'' provides a conceptually
clear approach to constructing a joint density, from which we
immediately have the Bayesian posterior
\begin{equation}
P(C_{\ell} | \Bd) = \int d\Bs \ P(C_{\ell}, \Bs | \Bd)
\end{equation}
The relevance of the above for this paper lies in relating what we
refer to as the {\it joint posterior}, $P(C_{\ell}, \Bs | \Bd)$, and
the more familiar likelihood $\mathcal{L}(\Bd | C_{\ell}) \propto
P(C_{\ell} | \Bd) / P(C_{\ell})$,

Although we can analytically compute the integral of the joint
posterior over the signal for the Gaussian signal and noise processes
considered here, and therefore simply write down the functional form
of the likelihood, it is too expensive to evaluate it for any
specified $C_{l}$ given high-resolution data. Furthermore, for more
complicated data models (i.e. including foreground model
uncertainties) we will not be able to perform the integrals over the
additional degrees of freedom.  Both situations then instead motivate
sampling from the joint posterior, and thereby generating samples from
$P(C_{\ell} | \Bd)$ without ever evaluating $P(C_{\ell} | \Bd)$.  We
now discuss the original Gibbs sampling approach proposed and
implemented by \citet{jewell:2004}, \citet{wandelt:2004} and
\citet{eriksen:2004}, and then introduce a new MCMC step which
directly addresses the previously reported slow probabilistic
convergence in the low signal to noise regime \citep{eriksen:2004}.

\subsection{The CMB Gibbs sampler}
\label{sec:cmb_sampling}

As stated above, our goal is to sample from the joint posterior,
\begin{equation}
  - 2  \log P(\Bs, C_{\ell}|\Bd) =
  \chi^{2}(\Bd, \Bs) + 
  \Bs^{t} \BS^{-1} \Bs + \log | \BS| 
  + \log P(C_{\ell}).
\label{eq:cmb_posterior}
\end{equation}
For notational convenience, we have here dropped constant factors of
$2\pi$, and also defined
\begin{equation}
\chi^{2}(\Bs, \Bd) = (\Bd - \Bs)^{t} \BN^{-1}(\Bd -\Bs).
\end{equation}
One approach to sample from this posterior is to use an algorithm
known as Gibbs sampling, where we can alternately sample from the
respective conditional densities,
\begin{align}
\Bs^{i+1} &\leftarrow P(\Bs | C_{\ell}^i, \Bd) \\
C_{\ell}^{i+1} &\leftarrow P(C_{\ell} | \Bs^{i+1}, \Bd).
\end{align}
Here $\leftarrow$ indicates sampling from the distribution on the
right-hand side. After some burn-in period, during which all samples
must be discarded, the joint samples $(\Bs^i, C_{\ell}^i)$ will be
drawn from the desired density. Thus, the problem is reduced to that
of sampling from the two \emph{conditional} densities $P(\Bs |
C_{\ell}, \Bd)$ and $P(C_{\ell} | \Bs, \Bd)$.

We now describe the sampling algorithms for each of these two
conditional distributions, starting with $P(C_{\ell} | \Bs, \Bd)$.
First, note that $P(C_{\ell} | \Bs, \Bd) = P(C_{\ell} | \Bs)$ which
follows directly from the construction of the joint density of
``everything'' above.  This is also intuitively easy to understand
since if we already know the CMB sky signal, the data themselves tell
us nothing new about the CMB power spectrum. Next, since the sky is
assumed to be Gaussian and isotropic, the distribution reads
\begin{equation}
P(C_{\ell} | \Bs) \propto P(C_{\ell}) \frac{e^{-\frac{1}{2}
    \Bs_{\ell}^{t}\BS_{\ell}^{-1}\Bs_{\ell}}}{\sqrt{|\BS_{\ell}|}} =    
P(C_{\ell})
\frac{e^{-\frac{2\ell+1}{2} \frac{\sigma_{\ell}}{C_{\ell}}}}{C_{\ell}^{\frac{2\ell+1}{2}}},
\end{equation}
which, when interpreted as a function of $C_{\ell}$, is known as the
inverse Gamma distribution. In this expression, $\sigma_{\ell} =
\frac{1}{2\ell+1} \sum_{m} |a_{\ell m}|^2$ denotes the observed power spectrum
of $\Bs$. Fortunately, there exists a simple textbook sampling
algorithm for this distribution \citep[e.g.,][]{gupta:2000}, and we
refer the interested reader to the previous papers for details. For an
alternative, and more flexible, sampling algorithm, see
\citet{wehus:2008}.

In order to describe the sky signal sampling step, we first define the
mean-field map (or Wiener filtered data) to be $\hat{\Bs} = (\BS^{-1}
+ \BN^{-1})^{-1} \BN^{-1} \Bd$, and note that the conditional sky
signal density given the data and $C_{l}$ can be written as
\begin{align}
P(\Bs | C_{\ell}, \Bd) &\propto e^{-\frac{1}{2} (\Bs-\hat{\Bs})^t (\BS^{-1} + \BN^{-1}) (\Bs-\hat{\Bs})}.
\end{align}
Thus, $P(\Bs | C_{\ell}, \Bd)$ is a Gaussian distribution with mean
equals to $\hat{\Bs}$ and a covariance matrix equals to $(\BS^{-1} +
\BN^{-1})^{-1}$.

Sampling from this Gaussian distribution is straightforward, but
computationally somewhat cumbersome. First, draw two random white
noise maps $\omega_0$ and $\omega_1$ with zero mean and unit
variance. Then solve the equation
\begin{equation}
\left[\BS^{-1} + \BN^{-1}\right] \Bs = \BN^{-1}\Bd + \BS^{-\frac{1}{2}} \omega_0 +
\BN^{-\frac{1}{2}} \omega_1.
\label{eq:lin_sys}
\end{equation}
for $\Bs$. Since the white noise maps have zero mean, one immediately
sees that $\langle \Bs \rangle = \hat{\Bs}$, while a few more
calculations show that $\langle \Bs \Bs^{t} \rangle = (\BS^{-1} +
\BN^{-1})^{-1}$. 

The problematic part about this sampling step is the solution of the
linear system in Equation \ref{eq:lin_sys}. Since this a $\sim10^6
\times 10^6$ system for current CMB data sets, it cannot be solved by
brute force. Instead, one must use a method called Conjugate Gradients
(CG), which only requires multiplication of the coefficient matrix on
the left-hand side, not inversion. For details on these computations,
together with some ideas on preconditioning, see \citet{eriksen:2004}.

\subsection{Convergence issues in the low signal-to-noise regime}

As originally applied to high-resolution CMB data, the Gibbs sampling
algorithm as described above has very slow convergence at the
high-$\ell$, low signal-to-noise part of the spectrum.  The reason for
the slow convergence is easy to understand in light of the above: When
sampling from $P(C_{\ell} | \Bs)$, the typical step size is given by
cosmic variance at all angular scales. In the high signal-to-noise
regime, cosmic variance dominates the noise variance, and we are able
to explore the full width of the posterior in only a few Gibbs
iterations. However, in the low signal-to-noise end, cosmic variance
is far smaller than the posterior variance, and it takes a
prohibitively long time to converge probabilistically.  This problem
of ``slow mixing'' of the Gibbs sampler is illustrated in figures
\ref{fig:TT_trace_plots} and \ref{fig:TT_correlation_length}.  The
long correlation length starting at signal-to-noise of unity leads to
extremely long run times in order to produce a reasonable number of
uncorrelated samples.

\section{A Low Signal-to-Noise MCMC Sampler}

When sampling from the true posterior, the goal is to produce as many
independent samples from $P(C_{\ell}, \Bs | \Bd)$ as possible.  One might
intuitively guess that it should be straightforward to establish good
approximations to the posterior in the low signal-to-noise regime,
since in the limit of vanishing signal to noise we simply recover the
prior. This suggests that we look for a sampling scheme in which we
first sample a new spectrum from some approximation to the true
posterior independent on the current spectrum and CMB map, followed by
sampling the CMB map from the conditional $P(\Bs|C_{\ell}, \Bd)$. The
problem with such a direct scheme is that the accept probability will
involve a ratio of determinants which are too expensive to compute.

We are therefore motivated to look for a sampling scheme in which we
can make a large variation in $C_{\ell}$ in the low signal-to-noise
regime, and make an associated {\it deterministic change} in the CMB
map, while still maintaining a reasonably high acceptance rate. The
motivation for a deterministic change is that it will avoid
introducing ratios of determinants which we cannot compute. 

\subsection{Proposal rule and acceptance probability}

Assume that we have defined a deterministic
sampling scheme for $\Bs$, and that our new CMB map is given by some
function
\begin{equation}
\Bs_{n+1} = F(\Bs_{n}, C_{\ell}^{(n+1)}, C_{\ell}^{(n)} ).
\end{equation}
Then the condition of detailed balance for our MCMC
sampler requires that
\begin{equation}
F^{-1}(\Bs_{n+1}, C_{\ell}^{(n+1)}, C_{\ell}^{(n)}) = F(\Bs_{n+1}, C_{\ell}^{(n)}, C_{\ell}^{(n+1)}),
\end{equation}
or, in other words, that the inverse function is given by exchanging
the order of the spectra in the function $F$. One simple function which has this property is
\begin{equation}
\Bs_{n+1} = \left(\frac{C_{\ell}^{(n+1)}}{C_{\ell}^{(n)}}\right)^{\frac{1}{2}} \Bs_{n}
\end{equation}
The total proposal matrix is then
\begin{eqnarray}
w(C_{\ell}^{(n+1)}, \Bs_{n+1} | C_{\ell}^{(n)}, \Bs_{n}) & = &  w(C_{\ell}^{(n+1)} | C_{\ell}^{(n)}, \Bd)
\nonumber \\
& &  \delta \left( \Bs_{n+1} - \left(\frac{C_{\ell}^{(n+1)}}{C_{\ell}^{(n)}}\right)^{-\frac{1}{2}} \Bs_{n} \right),
\nonumber
\end{eqnarray}
and the ``reverse'' proposal is
\begin{eqnarray}
  w(C_{\ell}^{(n)}, \Bs_{n} | C_{\ell}^{(n+1)}, \Bs_{n+1}) & = &  w(C_{\ell}^{(n)} | C_{\ell}^{(n+1)}, \Bd)
  \nonumber \\
  & &  \delta \left( \Bs_{n} - \left(\frac{C_{\ell}^{(n)}}{C_{\ell}^{(n+1)}}\right)^{-\frac{1}{2}} \Bs_{n+1} \right).
  \nonumber
\end{eqnarray}
The condition of detailed balance including deterministic moves
requires the consideration of some technical points which we leave
to Appendix \ref{app:proof}. There
we show that the full Metropolis-Hastings accept probability reads
\begin{eqnarray}
A & = & \min \left[ 1,
\frac{e^{- \chi^{2}(\Bs_{n+1}, \Bd)}}{e^{-\chi^{2}(\Bs_{n}, \Bd)}}
\frac{w(C_{\ell}^{(n)} | C_{\ell}^{(n+1)}, \Bd)}{w(C_{\ell}^{(n+1)} | C_{\ell}^{(n)}, \Bd)}
\right]
\end{eqnarray}
The significance of the above is that we can make relatively large
changes to the power spectrum in the low signal-to-noise regime, where
$\BN^{-1}$ is getting small, since the $\chi^2$ is affected only very
mildly by changes in any low signal-to-noise mode.

We note the interesting point (discussed more completely in Appendix
\ref{app:cov}) that if one changes variables in the
joint posterior from CMB maps, $\Bs$, to whitened maps, $\Bx =
\BC_{\ell}^{-\frac{1}{2}} \Bs$, and then Gibbs sample in the new
variables $(C_{\ell}, \Bx)$, the resulting accept probability is
numerically identitical to the above.  However, we note
the distinction here to emphasize the difference between MCMC
algorithms implementing deterministic proposals of maps given
$C_{\ell}$, and those sampling in a different set of variables, as
there could be other deterministic proposal schemes or
another change of variables which lead to improvements over the approach
presented in this paper.

For the numerical demonstration of the  MCMC algorithm presented
in this paper, we use a simple symmetric Gaussian proposal, truncated
at $C_{\ell}>0$ (or, for polarization, the region where the resulting
CMB covariance matrix is positive definite), for the power spectrum,
\begin{equation}
w(C_{\ell}^{(n+1)} | C_{\ell}^{(n)}, \Bd) \propto e^{-\frac{1}{2}
  \left(\frac{C_{\ell}^{n+1}-C_{\ell}^{n}}{\tau_{\ell}}\right)^2} I(C_{\ell}>0),
\end{equation}
where $\tau_{\ell}$ is a measure of the typical step size taken
between two samples. Note that because this proposal density is
symmetric, the ratio of $C_{\ell}$ proposals cancels, and the
acceptance probability is entirely determined by the change in
$\chi^{2}$.

It should be noted that while the above MCMC step satisfies detailed
balance, it is not {\it irreducable}, in the sense that there is not a
non-vanishing probability in reaching any state from any other state
in a finite number of MCMC steps; the phases are unchanged in each
MCMC step. However, alternating these steps with a traditional Gibbs
sampling step gives a combined ``two-step'' MCMC algorithm which
indeed is irreducable, and therefore provably converges to the joint
posterior. Once again, the details are left to the appendix for the
interested reader.

\subsection{Optimization of the MCMC sampler}

A general advantage of the Gibbs sampler is the fact that it is free
of tunable efficiency parameters. The same is not true for the
Metropolis-Hastings MCMC algorithm; for satisfactory sampling
performance, it typically has to be tuned quite extensively. In this
section, we describe three specific features that helps in this task,
namely 1) step size tuning, 2) slice sampling and 3) binning.

First, we have to ensure that the step size of our Gaussian proposal
density roughly matches the width of the target distribution, in order
to maintain both a reasonable acceptance rate and high mobility. We do
this by performing an initial test run, producing typically a few
hundreds $C_{\ell}$ samples, and compute the standard deviation of
these samples for each $\ell$. These are then adopted as the proposal
widths for the main run, scaled by some number less than unity,
typically between 0.05 and 0.5. For the initial test run, we
approximate the posterior width by the noise variance alone,
\begin{equation}
\tau_{\ell}^{2} = \frac{2}{2\ell+1} \frac{N_{\ell}}{b_{\ell}^{2}},
\end{equation}
because the MCMC sampler is used only in the low signal-to-noise
regime. In this expression $N_{\ell}$ is the power spectrum of the
instrumental noise alone, and $b_{\ell}$ is the product of the 
Legendre transform of the beam and the HEALPix window function.

Next, Metropolis-Hastings MCMC is inefficient in spaces with too many
free parameters. For this reason, we divide the power spectrum
coefficients, $C_{\ell}$, into subsets, each containing typically only
10--20 multipoles. Then we propose changes to one subset at a time,
while keeping all other multipoles fixed. Finally, we loop over
subsets, and thus effectively implement a multipole slice Gibbs
sampler for the full power spectrum.

This is computationally feasible, because a single MCMC proposal only
requires a single $\chi^{2}$ evaluation, which has a computational cost
of a single spherical harmonic transform. Since drawing a full sky map
from $P(\Bs|C_{\ell}, \Bd)$ in the classical Gibbs sampling step
requires $\mathcal{O}(10^{2})$ spherical harmonic transforms, we can
indeed afford to perform many MCMC proposals for each Gibbs step,
without dominating the total cost.

\begin{figure}
\mbox{\epsfig{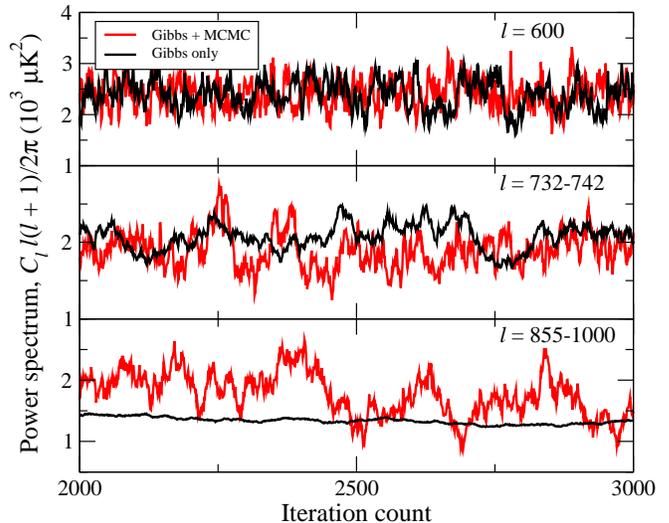}}
\caption{Comparison of $C_{\ell}$ chains produced by standard Gibbs
  sampling (black) and by the Gibbs+MCMC hybrid (red) for three
  selected multipole bins. The simulation was based on full sky
  coverage and uniform noise. See text for full details. }
\label{fig:TT_trace_plots}
\end{figure}

Nevertheless, for very high-resolution analysis it is often beneficial
to bin several $C_{\ell}$'s together, both in order to increase the
signal-to-noise of the joint coefficient, and to decrease the number
of parameters that needs to be sampled by MCMC. We implement this by
defining a new binned spectrum, weighted by $\ell(\ell+1)/2\pi$, as
follows,
\begin{equation}
C_{b} = \frac{1}{N_b}\sum_{\ell \in b} \frac{\ell(\ell+1)}{2\pi} C_{\ell}.
\end{equation}
Here $b=[\ell_{\textrm{min}}, \ell_{\textrm{max}}]$ denotes the
current bin, and $N_{b} = \ell_{\textrm{max}}-\ell_{\textrm{min}}+1$
is the number of multipoles within the bin. These new (and fewer)
coefficients are then sampled with the above MCMC sampler, after which
the original spectrum coefficients are given by
\begin{equation}
C_{\ell} = \frac{2\pi}{\ell(\ell+1)} C_{b}.
\end{equation}

\begin{figure}
\mbox{\epsfig{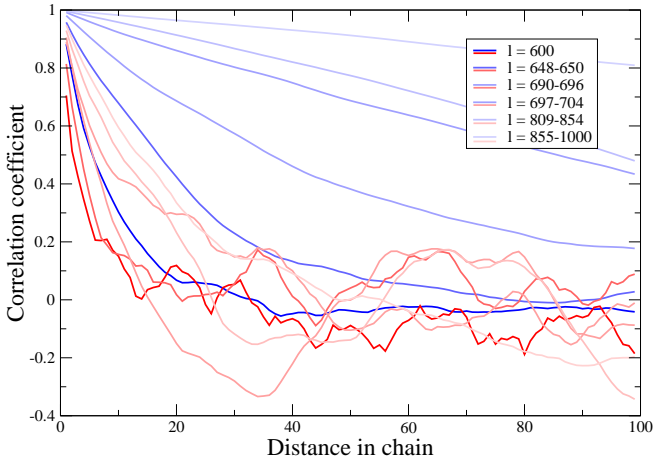}}
\caption{Comparison of chain correlation functions for standard Gibbs
  sampling (blue) and Gibbs+MCMC (red), computed from the full-sky
  uniform noise temperature data set.. Note that while the correlation
  length goes to infinity with increasing $\ell$ (or equivalently, low
  signal-to-noise) for standard Gibbs sampling, it is $\lesssim40$
  everywhere for the MCMC hybrid case. }
\label{fig:TT_correlation_length}
\end{figure}

\section{Testing and Validation}
\label{sec:simulations}

We have implemented the new sampling step described above in the
previously Gibbs sampling code called ``Commander''
\citep{eriksen:2004,eriksen:2008a}, and in this section we demonstrate
its advantages compared to the old sampling algorithm. We consider two
different cases, namely high-$\ell$ temperature and low-$\ell$
polarization analysis. In the former case, we also analyse two cases,
with and without a sky cut. The former allows us to verify the results
against an analytically known answer, while the second demonstrates
that the sky cut does not degrade the sampling efficiency.

\subsection{Temperature analysis}

The high-$\ell$ temperature simulation is designed to mimic the 5-year
WMAP temperature data \citep{hinshaw:2008} with one exception, namely
that the noise is assumed spatially uniform, in order to facilitate
analytic comparison. Specifically, the CMB realization was drawn from
the best-fit $\Lambda$CDM model derived from WMAP alone
\citep{komatsu:2008}, including multipoles up to
$\ell_{\textrm{max}}=1000$, and then smoothed with the instrumental
beam of the WMAP V1 differencing assembly, and pixelized at
HEALPix\footnote{http://healpix.jpl.nasa.gov} resolution
$N_{\textrm{side}}=512$. Finally, uniform noise of $\sigma_0 =
40\mu\textrm{K}$ RMS was added to each pixel. This corresponds to a
signal-to-noise ratio of unity at $\ell \sim 550$, roughly similar to
the 5-year WMAP data. We analyse this simulation both with and without
the WMAP KQ85 sky cut \citep{gold:2008}.

In both analyses, we adopted the Gaussian proposal density with tuned
variances, as described above. We also bin the power spectrum in
progressively wide bins, starting at $\ell = 600$, to maintain a
reasonable signal-to-noise per sampled power spectrum parameter. Ten
bins were sampled jointly per proposal, while all others were kept
fixed.

In the full-sky case, we produced a total of 31,800 samples over 60
chains, and in the cut sky case a total of 6800 samples. The cost for
producing one sample in the latter, and by far most expensive, set was
2.5 CPU hours, for a total of 17\,000 CPU hours. The number of MCMC
steps per Gibbs step was one in the former and 20 in the latter. 
(Since the the signal sampler dominates the cut sky Gibbs chain one can
perform more low S/N steps without slowing down the overall code significantly.)
In addition to these two main sample sets, we also
produced two longer chains with each 3500 samples for the full-sky
casee, both with and without the new MCMC step turned on, in order to
compare the Markov chain correlation lengths before and after
including the MCMC sampler.

We first consider the full-sky data set, and in Figure
\ref{fig:TT_trace_plots} we show a segment of each of the two longer
chains for three selected multipole bins. The top panel shows
$\ell=600$, which is the first bin to be sampled by MCMC, the middle
panel shows $\ell=732-742$, where there is still some signal in the
data, and, finally, the bottom panel shows $\ell=855-1000$, which is
strongly noise dominated. Starting with the top panel, we see that the
red curve (Gibbs+MCMC) scatters significantly faster than the black
curve (Gibbs only), implying more efficient sampling. This trend
becomes even stronger with lower signal-to-noise, until the last case,
where the Gibbs-only chain essentially does not move at all, while the
MCMC sampler does probe the full range. Note, however, that even the
MCMC sampler has a significant correlation length in this range, and
this implies that there is still some room for improvement to be made
in defining our proposals.

Next, these considerations are quantified in Figure
\ref{fig:TT_correlation_length}, where we plot the Markov chain
correlation length as a function of distance in the chain, for six
bins with and without the MCMC sampler. As first reported by
\citet{eriksen:2004}, we see that the Gibbs-only correlation length
increases dramatically with decreasing signal-to-noise, rendering the
algorithm essentially useless in this regime. However, we also see
that the new MCMC step effectively resolves this issue, as the
correlation length (here defined by having a correlation less than
0.2) now is less than $\sim40$ steps. This is a dramatic improvement,
and makes the algorithm useful even in this range. Nevertheless, we
once again point out that it is possible to make further improvements
by establishing better proposal densities.

\begin{figure}
\mbox{\epsfig{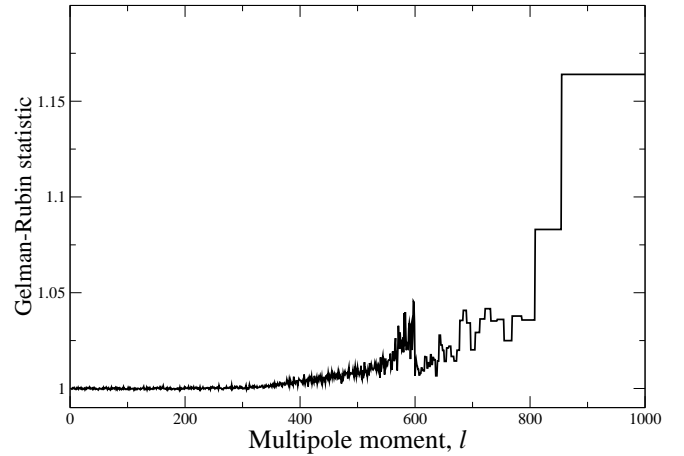}}
\caption{Gelman-Rubin statistic for the full-sky, uniform noise
  temperature analysis. Note the feature at $\ell=600$, which marks
  the transition between standard Gibbs sampling and Gibbs+MCMC.}
\label{fig:TT_gr}
\end{figure}

In Figure \ref{fig:TT_gr} we consider the convergence properties of
the $\sim30$k samples set, by computing the Gelman-Rubin statistic $R$
\citep{gelman:1992} as a function of $\ell$. Typically, one recommends
that $R$ should be less than, say, 1.2 in order to claim
convergence. We see that this holds everywhere for this sample set,
and typically it is even less than 1.05. Note also the step at
$\ell=600$, showing clearly the beneficial effect of the MCMC
sampler.

\begin{figure}
\mbox{\epsfig{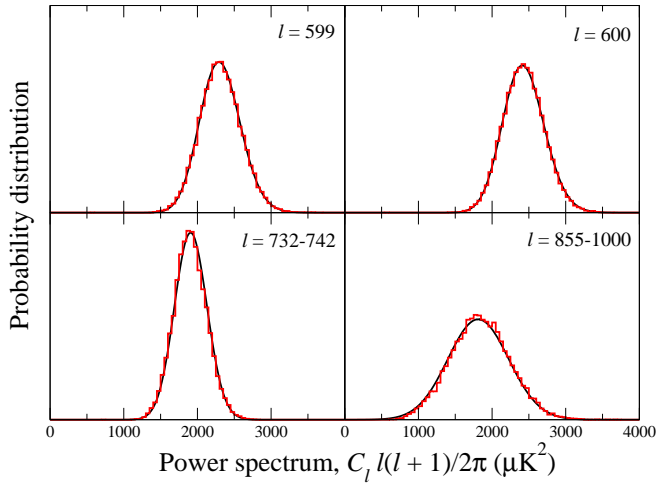}}
\caption{High-$\ell$ temperature marginal posteriors computed with
  Gibbs+MCMC from the full-sky, uniform noise temperature data set,
  compared to analytic results.}
\label{fig:TT_posteriors}
\end{figure}

Next, in Figure \ref{fig:TT_posteriors} we compare the marginal
distributions derived from this sample set with the analytic result,
\begin{equation}
P(C_{\ell}|\mathbf{d}) \propto \prod_{\ell \in b}
\frac{e^{-\frac{2\ell+1}{2}
    \frac{\sigma_{\ell}^{\textrm{S+N}}}{b_{\ell}^2
        C_{\ell}+N_{\ell}}}}
 {(b_{\ell}^2 C_{\ell}+N_{\ell})^{\frac{2\ell+1}{2}}}.
\end{equation}
Here $b=[\ell_{\textrm{min}}, \ell_{\textrm{max}}]$ indicates a given
multipole bin, $b_{\ell}$ denotes the product of the instrumental beam
and the HEALPix pixel window, and $\sigma_{\ell}^{\textrm{S+N}}$ is
the power spectrum of the noisy data map. We see that the new
algorithm reproduces the analytic distributions very well, and this
verifies the overall method.

\begin{figure}
\mbox{\epsfig{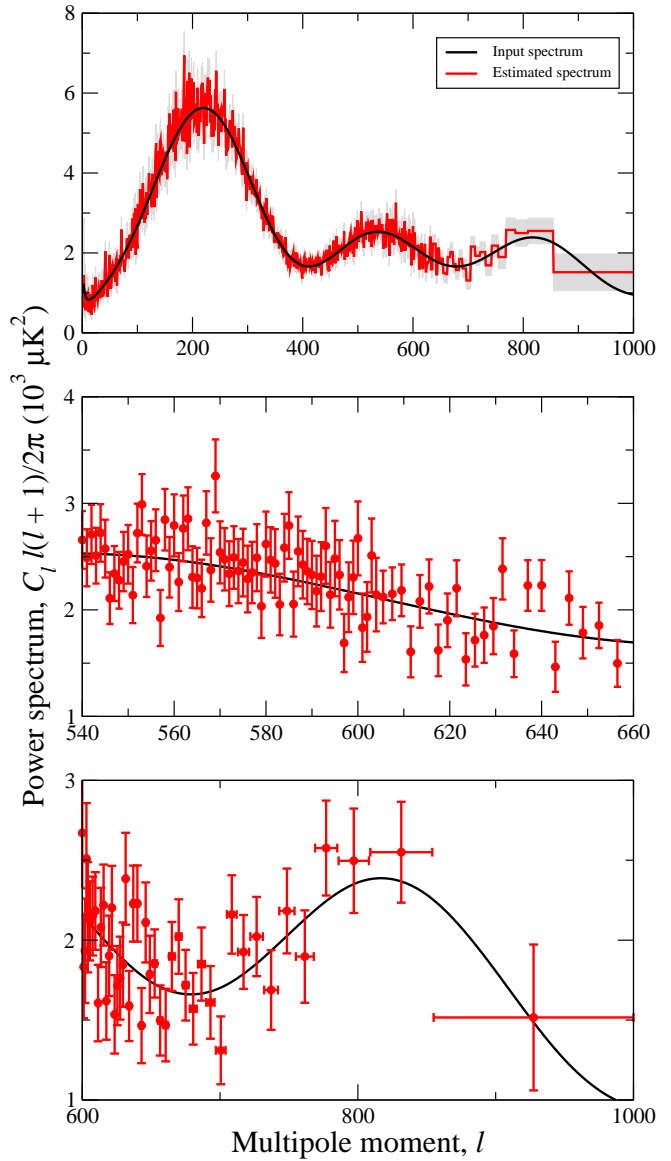}}
\caption{Temperature power spectrum estimated from cut sky temperature
data. The panels show the same spectrum, but emphasizing different
multipole ranges (full-range; S/N$\sim$1 transition region; and
high-$\ell$, low S/N).}
\label{fig:TT_spectrum}
\end{figure}

Finally, the cut-sky power spectrum with one-sigma confidence regions
is shown in three panels in Figure \ref{fig:TT_spectrum}, focusing on
different $\ell$-ranges, namely all $\ell$'s, the $S/N \sim 1$
transition region, and the low $S/N$ region. This completes the
high-$\ell$ temperature analysis validation.

\subsection{Polarization analysis}

We now consider polarization analysis, and construct a new low-$\ell$
simulation for this purpose. This simulation does not mimic any
planned experiment, but is rather designed to highlight the analysis
method itself. Specifically, we drew a new CMB realization from the
best-fit WMAP $\Lambda$CDM spectrum that includes a non-zero tensor
contribution, including multipoles up to $\ell_{\textrm{max}}=150$,
and convolved this with a $3^{\circ}$ FWHM Gaussian beam, and
pixelized it at $N_{\textrm{side}} = 64$. Uniform noise of
$5\mu\textrm{K}$ RMS was added to the temperature component, and
$1\mu\textrm{K}$ RMS to the polarization components. The 5-year WMAP
polarization sky mask was imposed on the data.

We allowed for non-zero $C_{\ell}^{TT}$, $C_{\ell}^{TE}$,
$C_{\ell}^{EE}$ and $C_{\ell}^{BB}$ spectra, but fixed $C_{\ell}^{TB}
= C_{\ell}^{EB} = 0$. These spectra were then individually binned to
maintain a reasonable signal-to-noise per bin. (Details on how to
introduce individual binning of each power spectrum were recently
described by Eriksen and Wehus, 2008.) Again, a tuned Gaussian proposal
density was used in the MCMC step. A total of 12\,000 samples were
produced over 12 chains, and the CPU time per sample was 55 seconds,
for a total of $\sim200$ CPU hours.

\begin{figure}
\mbox{\epsfig{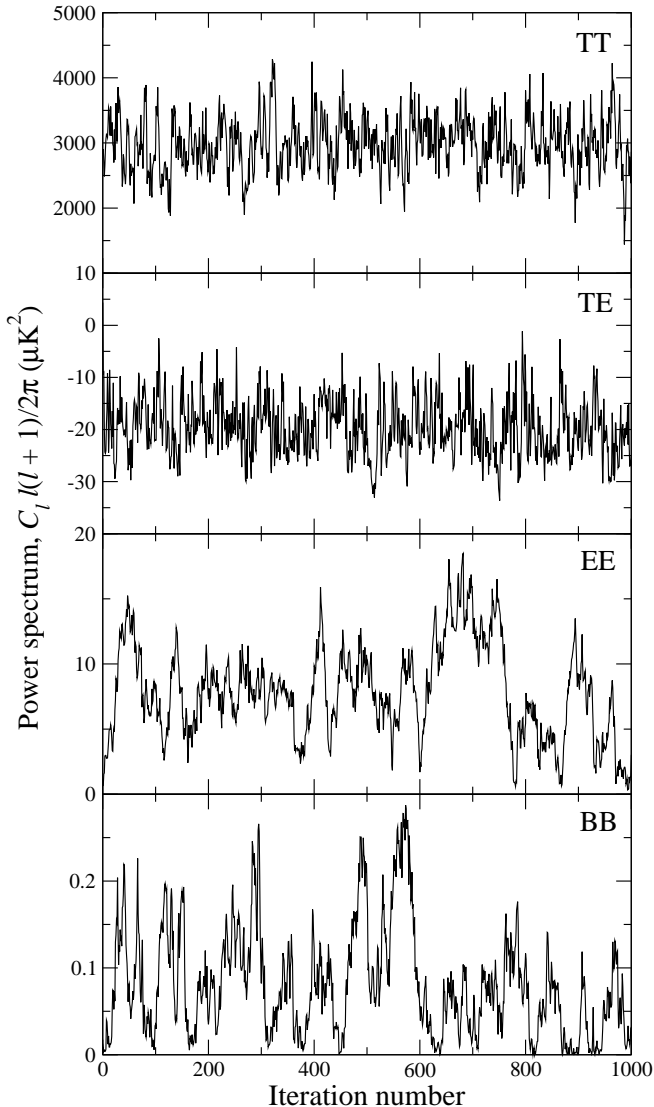}}
\caption{$C_{\ell}$ chains generated by Gibbs+MCMC hybrid for the
  cut-sky polarization data set. Only the highest multipole bin for
  each spectrum is shown ($\ell = 108-150$ for TT, $\ell = 88-150$ for
  TE, $\ell=101-150$ for EE and $\ell=61-150$ for BB).}
\label{fig:pol_trace_plots}
\end{figure}

In Figure \ref{fig:pol_trace_plots} we show one $C_{\ell}$ chain for
each of the four sampled spectra, for the last (and therefore most
difficult) bin in each case. Note that the $C_{\ell}^{EE}$ and
$C_{\ell}^{BB}$ spectra have essentially vanishing signal-to-noise,
and therefore these chains reach zero values. Clearly, we see that
mixing properties of these chains are satisfactory, and the
correlation lengths are quite short.

\begin{figure}
\mbox{\epsfig{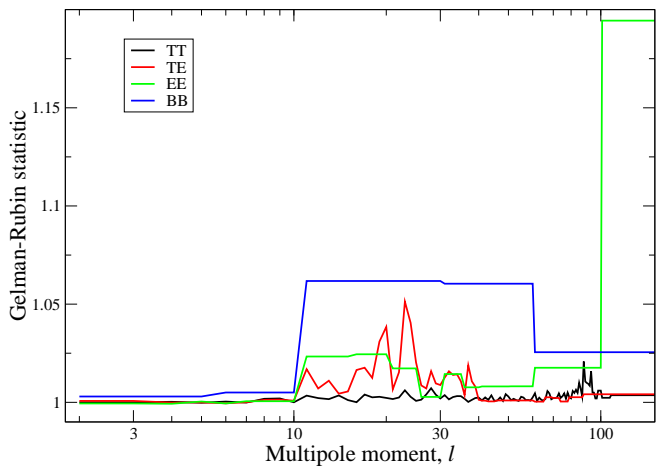}}
\caption{Gelman-Rubin statistic for cut-sky polarization analysis.}
\label{fig:gr_polarization}
\end{figure}

In Figure \ref{fig:gr_polarization} we show the Gelman-Rubin
statistics for each of the four power spectra, and with the single
exception of the very last bin of $C_{\ell}^{EE}$, all $R$ values are
well below 1.1. Thus, all spectra have converged well everywhere.

\begin{figure}
\mbox{\epsfig{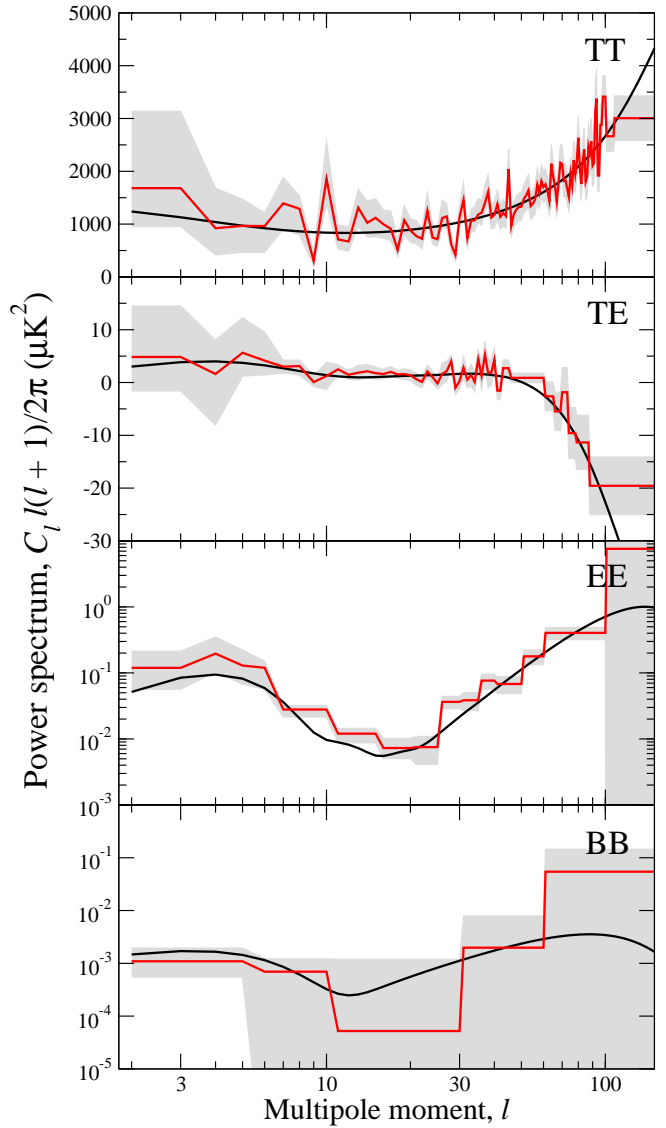}}
\caption{Marginal $C_{\ell}$ power spectra (red curves) estimated from
  cut sky polarization data. Gray bands indicate 68\% confidence
  regions, and the black lines show the input spectrum. (Note that the
  marginal spectra shown here are not individually unbiased estimators
  because of the correlations between TT, TE and EE. Proper treatment
  of the full joint polarization density will be considered separately
  in a future publication.)}
\label{fig:pol_spectrum}
\end{figure}

Finally, in Figure \ref{fig:pol_spectrum} we show the reconstructed
marginal power spectra for each polarization component, overplotted on
the input spectrum. The agreement is very good. Note, however, that
these spectra are direct marginals, and not a joint maximum likelihood
estimate. They are therefore not individual unbiased estimators. In
particular, the marginal $C_{\ell}^{EE}$ power spectrum is biased
slightly high because of the combination of the
$C_{\ell}^{TT}C_{\ell}^{EE} - (C_{\ell}^{TE})^2 > 0$ positivity
constraint and relatively low signal-to-noise. Consideration of the
joint polarization posterior, which \emph{is} an unbiased estimator,
is postponed to a future publication.



\section{Conclusions}
\label{sec:conclusions}

We have presented a new MCMC algorithm for the high-L, low
signal to noise limit of the joint posterior which
solves the slow probabilistic convergence of the traditional
Gibbs sampler in this regime.  This in principle allows sampling over the
joint posterior $p(C_{l}, \Bs | \Bd)$ over the entire range
of angular scales probed by current and future CMB experiments.
The limiting computational burden is now entirely in the map-making
step of Gibbs sampling, for which the cost per Gibbs iteration
now scales with the expense of multiplication by the inverse
noise matrix $\BN^{-1}$.  Assuming pixel uncorrelated (but scan weighted)
noise as a good approximation at small angular scales, the cost of
an $\BN^{-1}$ multiplcation is that of a forward and inverse spherical
harmonic transform, or ${\cal O}(\ell_{\textrm{max}}^{3})$.  Future work will attempt to push
the generalized Gibbs + MCMC sampling scheme presented here to smaller
angular scales, ultimately limited by the degree to which we can compute
harmonic transforms.

\begin{acknowledgements}
  We acknowledge use of the
  HEALPix\footnote{http://healpix.jpl.nasa.gov} software
  \citep{gorski:2005} and analysis package for deriving the results in
  this paper. HKE acknowledges financial support from the Research
  Council of Norway.
\end{acknowledgements}

\appendix

\section{Including Deterministic Proposals in MCMC}
\label{app:proof}
Here we review the derivation of the accept probability in Markov Chain Monte Carlo
when using deterministic proposals (or proposals where some of the degrees of freedom
are specified as deterministic functions of the past state and/or proposed
variations in some other degrees of freedom).  We first briefly review
the Metropolis-Hastings Markov Chain Monte Carlo algorithm and the proof of
its convergence, and then turn to the special case involving deterministic
proposals.  Much of the review of the MCMC algorithm here follows \citep{Sokal:1989}.
We also note that similar technical considerations including deterministic
elements in proposals are presented in \citep{Green:1995} in the context of
MCMC algorithms in which the dimension of the state space itself is included
as a random variable to be sampled over.

The goal is the construction of a transition matrix
$T(C_{l}, \Bs | C_{l}', \Bs', \Bd)$ such that after initializing
the Markov Chain with a sample from any probability density $p_{0}(C_{l}, \Bs | \Bd)$,
we generate samples from a sequence of probability densities
\begin{equation}
p_{n+1}(C_{l}, \Bs | \Bd)  \equiv   \int d(C_{l}', \Bs') \ 
T(C_{l}, \Bs | C_{l}', \Bs', \Bd) \ p_{n}(C_{l}', \Bs' | \Bd)
\end{equation}
which eventually converge to an {\it equilibrium density} $\pi(C_{l}, \Bs | \Bd)$
\begin{equation}
\pi (C_{l}, \Bs | \Bd) = \lim_{n \rightarrow \infty}  p_{n}(C_{l}, \Bs | \Bd)
\end{equation}
We remind the reader
of the sufficient conditions to establish convergence of an MCMC algorithm:
{\it stationarity}, which means that the MCMC transition matrix satisfies
\begin{equation}
\pi(C_{l}, \Bs | \Bd) = \int d(C_{l}', \Bs') \ T(C_{l}, \Bs | C_{l}', \Bs', \Bd) \ \pi(C_{l}', \Bs' | \Bd)
\end{equation}
and {\it irreducability}, which means that for any two states, there is a finite
number of iterations which give a non-vanishing probability to transition from one
state to the other.  It is well known that these two properties are sufficient
to establish convergence, as can be seen simply from the triangle inequality
\begin{eqnarray}
\int d(C_{l}, \Bs) \ \left| \pi(C_{l}, \Bs | \Bd) - p_{n}(C_{l}, \Bs | \Bd) \right|
& = & \int d(C_{l}, \Bs) \ \left| \int d(C_{l}', \Bs') \ T(C_{l}, \Bs | C_{l}', \Bs') 
\left( \pi(C_{l}', \Bs' | \Bd) - p_{n-1}(C_{l}', \Bs' | \Bd) \right) \right| \nonumber \\
& \le & \int d(C_{l}, \Bs) \  \int d(C_{l}', \Bs') \ T(C_{l}, \Bs | C_{l}', \Bs') 
\left| \pi(C_{l}', \Bs' | \Bd) - p_{n-1}(C_{l}', \Bs' | \Bd)  \right| \nonumber \\
& = & \int d(C_{l}', \Bs') \ \left( \int d(C_{l}, \Bs) \  T(C_{l}, \Bs | C_{l}', \Bs') \right) 
\left| \pi(C_{l}', \Bs' | \Bd) - p_{n-1}(C_{l}', \Bs' | \Bd)  \right| \nonumber \\
& = & \int d(C_{l}', \Bs') \ 
\left| \pi(C_{l}', \Bs' | \Bd) - p_{n-1}(C_{l}', \Bs' | \Bd)  \right| \nonumber
\end{eqnarray}

The Metropolis-Hastings Markov Chain Monte Carlo algorithm is one
method of constructing such a transition matrix.  We choose {\it any}
proposal matrix $w(C_{l}, \Bs | C_{l}', \Bs', \Bd)$ and then accept
the proposed move with a probability
\begin{equation}
0 \le A(C_{l}, \Bs | C_{l}', \Bs', \Bd) \le 1
\end{equation}
while rejecting the proposed move with probability $1 - A$ leads
to a ``null transition'' where the next state in the Markov Chain remains the same.
Application of this algorithm then leads to the sequence of probability densities
which satisfy
\begin{eqnarray}
p_{n+1}(C_{l}, \Bs | \Bd) & = & 
\left( 1 - \int d(C_{l}', \Bs') \ A(C_{l}', \Bs' | C_{l}
, \Bs, \Bd) w(C_{l}', \Bs' | C_{l}, \Bs, \Bd) \right)
p_{n}(C_{l}, \Bs | \Bd) \nonumber \\
& & +
\int d(C_{l}', \Bs') \ A(C_{l}, \Bs | C_{l}', \Bs', \Bd) w(C_{l}, \Bs | C_{l}', \Bs', \Bd)
p_{n}(C_{l}', \Bs' | \Bd) 
\end{eqnarray}
where the first term is the constribution to the probability density
$p_{n+1}$ if we reject any proposed move, while the second term
is the contribution from accepting the proposed move from any possible
previous state.  If we demand that, for a chosen proposal matrix, the accept probability satisfies
\begin{equation}
\pi(C_{l}', \Bs' | \Bd) w(C_{l}, \Bs | C_{l}', \Bs', \Bd) A(C_{l}, \Bs | C_{l}', \Bs', \Bd)
= A(C_{l}', \Bs' | C_{l}, \Bs, \Bd) w(C_{l}', \Bs' | C_{l}, \Bs, \Bd) \pi(C_{l}, \Bs | \Bd)
\end{equation}
then we see that the MH MCMC algorithm satisfies stationarity, i.e. denoting
by $T \circ \pi$ the density resulting from one application of the transition matrix
to $\pi$, we have directly from detailed balance
\begin{eqnarray}
T \circ \pi 
& = & \left( 1 - \int d(C_{l}', \Bs') \ A(C_{l}', \Bs' | C_{l}
, \Bs, \Bd) w(C_{l}', \Bs' | C_{l}, \Bs, \Bd) \right)
\pi(C_{l}, \Bs | \Bd) \nonumber \\ 
& & +
\int d(C_{l}', \Bs') \ A(C_{l}, \Bs | C_{l}', \Bs', \Bd) w(C_{l}, \Bs | C_{l}', \Bs', \Bd)
\pi(C_{l}', \Bs' | \Bd) \nonumber \\ 
& = & \left( 1 - \int d(C_{l}', \Bs') \ A(C_{l}', \Bs' | C_{l}
, \Bs, \Bd) w(C_{l}', \Bs' | C_{l}, \Bs, \Bd) \right)
\pi(C_{l}, \Bs | \Bd) \nonumber \\ 
& & + \pi (C_{l}, \Bs | \Bd)
\int d(C_{l}', \Bs') \ A(C_{l}', \Bs' | C_{l}, \Bs, \Bd) w(C_{l}', \Bs' | C_{l}, \Bs, \Bd)
 \nonumber \\ 
& = & \pi(C_{l}, \Bs | \Bd)
\end{eqnarray}

We now turn to the case where our proposal is of the form
\begin{equation}
w(\Bs', C_{l}' | \Bs, C_{l}) = \delta \left[ \Bs' - F(\Bs, C_{l}', C_{l}) \right] w(C_{l}' | C_{l}, \Bd)
\end{equation}
where we randomly propose a new power spectrum, posibly in a manner conditionally
denpendent on the current spectrum and the data, and then deterministically
compute a new CMB map with some function
\begin{equation}
\Bs' = F(\Bs, C_{l}', C_{l})
\end{equation}
To satisfy detailed balance with a non-vanishing accept probability
our function must satisfy
\begin{eqnarray}
\Bs' & = & F(\Bs, C_{l}', C_{l}) \nonumber \\
\Bs & = & F(\Bs', C_{l}, C_{l}')
\end{eqnarray}
or, that the inverse function is equivalent to interchanging the order
of the power spectrum arguements
\begin{equation}
 F(\Bs', C_{l}, C_{l}') = F^{-1}(\Bs',C_{l}', C_{l})
\end{equation}
In this paper, we have chosen one such function, given by
\begin{equation}
F(\Bs, C_{l}', C_{l}) = [\BC']^{1/2} [\BC]^{-1/2} \Bs
\end{equation}
where interchanging the spectra in the function above does in fact give
the inverse function itself.

Our job now is to {\it derive} the accept probability such that we
satisfy stationarity (as discussed above).  For the proposal with deterministic
changes to some of the degrees of freedom, stationarity is satisfied if
\begin{eqnarray}
(T \circ \pi)(C_{l}, \Bs | \Bd)
& = & \left[ 1 - \int d(C_{l}'', \Bs'') \ A[C_{l}'', \Bs'' | C_{l}, \Bs]
\delta[\Bs'' - F(\Bs, C_{l}'', C_{l})] w(C_{l}'' | C_{l}, \Bd) \right] \pi(C_{l}, \Bs | \Bd) \nonumber \\
& & + \int d(C_{l}', \Bs') \ A[\Bs, C_{l} | \Bs', C_{l}'] \delta[\Bs - F(\Bs', C_{l}, C_{l}')]
w(C_{l} | C_{l}' , \Bd) \pi(C_{l}', \Bs' | \Bd) \nonumber 
\end{eqnarray}
In order to determine the integral over the $\delta$-function in the
accept term above, we recall the identity for $\delta[G(\Bx)]$, where $G(\Ba) = 0$,
\begin{equation}
\delta[G(\Bx)] = \frac{\delta(\Bx - \Ba)}{\left| \partial G / \partial \Bx \right|_{a} }
\end{equation}
In our case, we can identify
\begin{equation}
G(\Bs') = \Bs - F(\Bs', C_{l}, C_{l}')
\end{equation}
which vanishes at $F^{-1}(\Bs,C_{l}, C_{l}') = F(\Bs, C_{l}', C_{l})$.  We also have the Jacobian
\begin{equation}
\left| \frac{\partial G}{ \partial \Bs'} \right|_{\Bs' = F^{-1}(\Bs, C_{l}, C_{l}')} 
= \left| \frac{\partial F}{ \partial \Bs'} \right|_{\Bs' = F^{-1}(\Bs, C_{l}, C_{l}')} 
\end{equation}
(i.e. $G(\Bs')$ is considered a function of $\Bs'$ with the other CMB map
$\Bs$ considered fixed) which therefore gives
\begin{equation}
\delta[\Bs - F(\Bs', C_{l}, C_{l}')] = 
 \delta[\Bs' - F^{-1}(\Bs, C_{l}, C_{l}')]
\left| \frac{\partial F}{ \partial \Bs'} \right|^{-1}_{\Bs' = F^{-1}(\Bs, C_{l}, C_{l}')} 
\end{equation}
Inserting this into the condition for stationarity we have
\begin{eqnarray}
(T \circ \pi )(C_{l}, \Bs | \Bd)
& = & \left[ 1 - \int d(C_{l}'', \Bs'') \ A[C_{l}'', \Bs'' | C_{l}, \Bs]
\delta[\Bs'' - F(\Bs, C_{l}'', C_{l})] w(C_{l}'' | C_{l}, \Bd) \right] \pi(C_{l}, \Bs | \Bd) \nonumber \\
& & + \int d(C_{l}', s') \ A[s, C_{l} | s', C_{l}'] 
\left( \delta[\Bs' - F^{-1}(\Bs, C_{l}, C_{l}')]
\left| \frac{\partial F}{ \partial \Bs'} \right|^{-1}_{\Bs' = F^{-1}(\Bs, C_{l}, C_{l}')} \right)
w(C_{l} | C_{l}' , \Bd) \pi(C_{l}', \Bs' | \Bd) \nonumber \\
& = & \left[ 1 - \int d(C_{l}'', \Bs'') \ A[C_{l}'', \Bs'' | C_{l}, \Bs]
\delta[\Bs'' - F(\Bs, C_{l}'', C_{l})] w(C_{l}'' | C_{l}, \Bd) \right] \pi(C_{l}, \Bs | \Bd) \nonumber \\
& & + \int d(C_{l}', \Bs') \ A[\Bs, C_{l} | \Bs', C_{l}'] 
\left( \delta[\Bs' - F(\Bs, C_{l}', C_{l})]
\left| \frac{\partial F}{ \partial \Bs'} \right|^{-1}_{\Bs' = F^{-1}(\Bs, C_{l}, C_{l}')} \right)
w(C_{l} | C_{l}' , \Bd) \pi(C_{l}', \Bs' | \Bd) \nonumber 
\end{eqnarray}
where in the second line we again used the property that the inverse $F^{-1}$ is equivalent
to $F$ with the spectra arguements interchanged.
We see from the above that a sufficient condition for stationarity is
\begin{equation}
\pi(C_{l}, \Bs | \Bd) w(C_{l}' | C_{l} , \Bd) 
A[\Bs', C_{l}' | \Bs, C_{l}] 
= A[\Bs, C_{l} | \Bs', C_{l}'] 
\left(  \left| \frac{\partial F}{ \partial \Bs'} \right|^{-1}_{\Bs' = F^{-1}(\Bs, C_{l}, C_{l}')} \right)
w(C_{l} | C_{l}' , \Bd) \pi(C_{l}', \Bs' | \Bd)
\end{equation}
An accept probability which satisfies this condition therefore gives cancellation
of the integrals over the $\delta$-functions for both the reject and accept
contributions, leaving us exactly with $T \circ  \pi = \pi$.
We therefore have the accept probability
\begin{equation}
A[\Bs', C_{l}' | \Bs, C_{l}] = \min \left[ 1,
\frac{\pi(C_{l}', \Bs' | d)}{\pi(C_{l}, \Bs | \Bd)}
\frac{w(C_{l} | C_{l}', \Bd)}{w(C_{l}' | C_{l}, \Bd)} 
\left(  \left| \frac{\partial F}{ \partial \Bs'} \right|^{-1}_{\Bs' = F^{-1}(\Bs, C_{l}, C_{l}')} \right)
\right]
\end{equation}
We give the expression above for the general case of any deterministic
change in the CMB map with a function which satisfies $F(\Bs, C_{l}, C_{l}') = F^{-1}(\Bs, C_{l}', C_{l})$.
We now explicitly evaluate this accept probability for the functional form chosen for this
paper.

Since we have $F(\Bs', C_{l}, C_{l}') = [\BC]^{1/2} [\BC']^{-1/2} \Bs'$, we have
\begin{equation}
\left(  \left| \frac{\partial F}{ \partial \Bs'} \right|_{\Bs' = F^{-1}(\Bs, C_{l}, C_{l}')} \right)
= \frac{|\BC|^{1/2}}{|\BC'|^{1/2}}
\end{equation}
Reminding the reader of the functional form of the joint posterior in
eqn. \ref{eq:cmb_posterior}, we have the accept probability given by
\begin{eqnarray}
A[\Bs', C_{l}' | \Bs, C_{l}] 
& = &  \min \left[ 1,
\frac{\pi(C_{l}', \Bs' | \Bd)}{\pi(C_{l}, \Bs | \Bd)}
\frac{w(C_{l} | C_{l}', \Bd)}{w(C_{l}' | C_{l}, \Bd)} 
\left(  \left| \frac{\partial F}{ \partial \Bs'} \right|^{-1}_{\Bs' = F^{-1}(\Bs, C_{l}, C_{l}')} \right)
\right] \nonumber \\
& = &  \min \left[ 1,
\frac{e^{- \chi^{2}(\Bs', \Bd)}}{e^{- \chi^{2}(\Bs, \Bd)}}
\frac{e^{- \Bs' [\BC']^{-1} \Bs'}}{e^{\Bs \BC^{-1} \Bs}}
\frac{|\BC|^{1/2}}{|\BC'|^{1/2}}
\frac{w(C_{l} | C_{l}', d)}{w(C_{l}' | C_{l}, d)} 
\left(  \left| \frac{\partial F}{ \partial \Bs'} \right|^{-1}_{\Bs' = F^{-1}(\Bs, C_{l}, C_{l}')} \right)
\right] \nonumber \\ 
& = &  \min \left[ 1,
\frac{e^{- \chi^{2}(\Bs', \Bd)}}{e^{- \chi^{2}(\Bs, \Bd)}}
\frac{e^{- \Bs' [\BC']^{-1} \Bs'}}{e^{\Bs \BC^{-1} \Bs}}
\frac{|\BC|^{1/2}}{|\BC'|^{1/2}}
\frac{w(C_{l} | C_{l}', \Bd)}{w(C_{l}' | C_{l}, \Bd)} 
\left(  \frac{|\BC'|^{1/2}}{|\BC|^{1/2}} \right) 
\right] \nonumber \\ 
& = &  \min \left[ 1,
\frac{e^{- \chi^{2}(\Bs', \Bd)}}{e^{- \chi^{2}(\Bs, \Bd)}}
\frac{w(C_{l} | C_{l}', \Bd)}{w(C_{l}' | C_{l}, \Bd)} 
\right] 
\label{eq:accept_prob}
\end{eqnarray}
where the last line follows from the invariance of the quadratic form
under the functional mapping
$\Bs' [\BC']^{-1} \Bs'  =  \Bs \BC^{-1} \Bs$.  Finally, we note that for the special
case of a symmmetric proposal matrix where $w(C_{l}' | C_{l}, \Bd) = w(C_{l} | C_{l}', \Bd)$, the
accept probability is completely determined by the (exponeniated) change in $\chi^{2}$
\begin{equation}
A[\Bs', C_{l}' | \Bs, C_{l}] =
  \min \left[ 1,
\frac{e^{- \chi^{2}(\Bs', \Bd)}}{e^{- \chi^{2}(\Bs, \Bd)}} \right] 
\end{equation}
As emphasized earlier in the main part of the text, the above allows large
changes to the spectrum precisely where the signal to noise is getting small,
as $\chi^{2}$ does not change much in this regime.

\section{Relation to Gibbs Sampling in a Change of Variables}
\label{app:cov}
We note here another interesting approach to an MCMC algorithm in a
{\it different set of variables} which in fact allows for large
moves in the spectrum in the low signal to noise regime.  We define the
CMB map
\begin{equation}
\Bx = \BC^{-1/2} \Bs
\end{equation}
We therefore have the joint posterior {\it in the new variables} according to
\begin{equation}
p(C_{l}, \Bs | \Bd) d(C_{l}, \Bs) = 
p(C_{l}, \Bx | \Bd) \left| \frac{\partial \Bs}{\partial x} \right| d(C_{l}, \Bx)
\end{equation}
which is explicitly, up to a normalization constant
\begin{equation}
-2 \log p(C_{l}, \Bx | d) = (\Bd - \BC^{1/2} \Bx) \BN^{-1}
(\Bd - \BC^{1/2} \Bx) - \| \Bx \|^{2}
\end{equation}
Then {\it traditional Gibbs sampling in the new variables} leads to an accept
probability when changing the spectrum given the change of variable map $x$ as
\begin{equation}
A(C_{l}', \Bx | C_{l}, \Bx) =
\min \left[ 1,
\frac{e^{-(\Bd - [\BC']^{1/2} \Bx) \BN^{-1}(\Bd - [\BC']^{1/2} \Bx)}}
{e^{-(\Bd - \BC^{1/2} \Bx) \BN^{-1}(\Bd - \BC^{1/2} \Bx)}}
\frac{w(C_{l} | \Bx, C_{l}', \Bd)}{w(C_{l}' | \Bx, C_{l}, \Bd)} \right]
\end{equation}
where in the above the proposed variation in the spectrum can now be
conditionally dependent on the current change of variable map $\Bx$.
Assuming a symmetric proposal, or one conditionally independent of $\Bx$
leads to an accept probability which is {\it numerically the same as 
\ref{eq:accept_prob} }, and also has the same property - large moves in the
spectrum are possible in the low signal to noise regime.
As a side note, we can see that $\log p(C_{l} | \Bx, \Bd)$
is quadratic in $C_{l}^{1/2}$, and suggests a proposal
given by a Gaussian in $C_{l}^{1/2}$.  However there are two problems with this
scheme - sampling in $C_{l}^{1/2}$ will result in re-introducing a Jacobian
factor given by the ratio of $|C'|^{1/2} / |C|^{1/2}$ which results typically
in low acceptance probabilities, and furthermore we cannot afford to exactly
compute the local ``Fisher'' covariance matrix for each $\Bx$.
Because of these difficulties, we in general need to produce a proposal
for $C_{l}$ and then compute the accept probability above.

We emphasize an important distinction between MCMC with deterministic
steps {\it in the original variables} $(C_{l}, \Bs)$ and Gibbs sampling
in the change of variables $(C_{l}, \Bx)$.  It is only for the specific
functional form that we have chosen for this paper that the numerical value of the
accept probabilities for $A(C_{l}' , \Bs' | C_{l}, \Bs)$ and $A(C_{l}', \Bx | C_{l}, \Bx)$
are the same.

At first glance, it might appear
that a random variation in some of the variables followed by a deterministic
change in the complementary set is always equivalent to random variation
in a new set of variables.  For notational convenience, we will assume the state
space is separated into two sets of variables $(\Bx,\By)$, i.e. for the CMB sampling
context we have $(\Bs, C_{l})$.  Now, to make the distinction between a change
of variables and deterministic steps in MCMC more precise, consider a ``global'' change of variables
of the form
\begin{eqnarray}
\Bu & = & F(\Bx,\By) \nonumber \\
\Bv & = & \By
\end{eqnarray}
with Jacobian
\begin{equation}
\left| \begin{array}{cc}
\frac{\partial \Bu}{\partial \Bx} & \frac{\partial \Bv}{\partial \Bx} \\
\frac{\partial \Bu}{\partial \By} & \frac{\partial \Bv}{\partial \By} \end{array}
\right| =
\left| \begin{array}{cc}
\frac{\partial F}{\partial \Bx} & 0 \\
\frac{\partial F}{\partial \By} & \id \end{array}
\right| = \left| \frac{\partial F}{\partial \Bx} \right|
\end{equation}
A Gibbs sampling step varying $v$ with $u$ fixed, has accept probability
\begin{eqnarray}
A(\By_{n+1}, \Bu_{n} | \By_{n}, \Bu_{n}) 
& = & \min \left[ 1,
\frac{\pi(\By_{n+1} | \Bu_{n}, \Bd)}{\pi(\By_{n} | \Bu_{n}, \Bd)}
\frac{w(\By_{n} | \Bu_{n}, \Bd)}{w(\By_{n+1} | \Bu_{n}, \Bd)} \right] \nonumber \\
& = & \min \left[ 1,
\frac{\pi(\By_{n+1}, \Bx_{n+1} | \Bd)}{\pi(\By_{n} , \Bx_{n} | \Bd)}
\left( \left| \frac{\partial F}{\partial \Bx} \right|_{\Bx_{n+1}, \By_{n+1}}
\left| \frac{\partial F}{\partial \Bx} \right|^{-1}_{\Bx_{n}, \By_{n}} \right)
\frac{w(\By_{n} | \Bu_{n}, \Bd)}{w(\By_{n+1} | \Bu_{n}, \Bd)} \right] \nonumber \\
\end{eqnarray}
where in the above we have the constraint
\begin{eqnarray}
\Bx_{n+1} & = & F^{-1}(\Bu_{n}, \By_{n+1}) \nonumber \\
\Bx_{n} & = & F^{-1}(\Bu_{n}, \By_{n}) 
\end{eqnarray}
Now consider an MCMC step in the original variables of the form
\begin{equation} 
w(\Bx_{n+1}, \By_{n+1} | \Bx_{n}, \By_{n }) = w(\By_{n+1} | \By_{n}, \Bx_{n}, \Bd)
\delta \left( x_{n+1} - H(x_{n}, y_{n+1}, y_{n}) \right)
\end{equation}
with general accept probability, according to the discussion above
\begin{eqnarray}
A(\By_{n+1}, \Bx_{n+1} | \By_{n}, \Bx_{n}) & = & \min \left[ 1,
\frac{\pi(\By_{n+1}, \Bx_{n+1} | \Bd)}{\pi(\By_{n}, \Bx_{n} | \Bd)}
\frac{w(\By_{n} | \Bx_{n}, \By_{n+1}, \Bd)}{w(\By_{n+1} | \Bx_{n+1}, \By_{n}, \Bd)}
\left( \left| \frac{\partial H}{\partial \Bx} \right|^{-1}_{\Bx_{n+1} = H^{-1}(\Bx_{n},\By_{n}, \By_{n+1})} \right)
\right]
\end{eqnarray}
Interestingly enough this suggests that we can set $H$ to be the function
\begin{equation}
H(\Bx, \By_{n+1}, \By_{n}) = F^{-1} \left( F(\Bx, \By_{n}), \By_{n+1} \right)
\end{equation}
Does this function have the correct properties for its inverse?  Assuming we
have computed in the forward direction $\Bx' = H(\Bx, \By_{n+1}, \By_{n})$, we can invert to find $x$ by
computing sequentially
\begin{eqnarray}
F(\Bx', \By_{n+1}) & = & F(\Bx, \By_{n}) \nonumber \\
\Bx & = & F^{-1} \left( F(\Bx', \By_{n+1}), \By_{n} \right) \nonumber \\
& \equiv & H(\Bx', \By_{n}, \By_{n+1})
\end{eqnarray}
where the last line follows from definition of the forward $H$.
Since we have, by definition
\begin{eqnarray}
\Bx' & = & H(\Bx, \By_{n+1}, \By_{n}) \nonumber \\
\Bx & \equiv & H^{-1}(\Bx', \By_{n+1}, \By_{n}) \nonumber 
\end{eqnarray}
we therefore have shown that
\begin{equation}
H^{-1}(\Bx', \By_{n+1}, \By_{n}) = H(\Bx', \By_{n}, \By_{n+1})
\end{equation}
as required for a non-vanishing accept probability.
The above as a function of $x$ has Jacobian
\begin{eqnarray}
\left| \frac{\partial H}{\partial \Bx} \right| 
& = & \left| \frac{\partial F^{-1}}{\partial \Bu}  \right|_{( \Bu(\Bx, \By_{n}), \By_{n+1})}
\  \left| \frac{\partial F}{\partial \Bx}  \right|_{(\Bx,\By_{n})} \nonumber \\
& = & \left| \frac{\partial F}{\partial \Bx}  \right|^{-1}_{( \Bx, \By_{n+1})}
\  \left| \frac{\partial F}{\partial \Bx}  \right|_{(\Bx,\By_{n})} \nonumber \\
\end{eqnarray}
However, when evaluated at $\Bx_{n+1} = H^{-1}(\Bx_{n}, \By_{n}, \By_{n+1})$, we will not
in general satisfy the required equality required for numerical equivalence
\begin{equation}
\left( \left| \frac{\partial H}{\partial \Bx} \right|^{-1}_{\Bx_{n+1} = H^{-1}(\Bx_{n},\By_{n}, \By_{n+1})} \right)
\neq  \left( \left| \frac{\partial F}{\partial \Bx} \right|_{\Bx_{n+1}, \By_{n+1}}
\left| \frac{\partial F}{\partial \Bx} \right|^{-1}_{\Bx_{n}, \By_{n}} \right)
\label{eq:jacobian_identity}
\end{equation}
So in general, while we can use any function $F(\Bx,\By)$ to generate deterministic
moves in the original variables within MCMC, this is not equivalent to
a Gibbs sampling step $p(\By_{n+1} | \Bu_{n}, \Bd)$ in the new variables using $(F(\Bx,\By), \By)$ as a global change of variables.

However, using the above construction for
the CMB change of variables, we have explcitly
\begin{eqnarray}
F^{-1} \left( F(\Bs, C_{l}), C_{l}' \right) & = & 
[\BC']^{1/2} \left( \BC^{-1/2} \Bs \right)
\end{eqnarray}
which is exactly the functional form used for the deterministic MCMC steps.  In this
case, it is because the Jacobian of our deterministic change in the
CMB map is independent of the current CMB map $\Bs$ (and only dependent
on the proposed and current spectra) that we have numerical
equivalence of the accept probabilities.

So in summary, while we can use any mapping $F(\Bx,\By)$ to generate deterministic
steps for use in MCMC, the accept probability is not equivalent to a conditional
step $p(\By | \Bu, \Bd)$ using $F(\Bx,\By)$ in a change of variables due to the general
``location'' dependence of the Jacobian.  Furthermore, setting
$H(\Bx, \By', \By) = F^{-1} ( F(\Bx,\By), \By')$ is not the most general form
for a function that satisfies the detailed balance requirement
$H^{-1}(\Bx, \By', \By) = H(\Bx, \By, \By')$.
In this sense then, a change of variables as an approach to more efficiently generating
samples from a probability density is distinct from a strategy of designing an MCMC
algorithm (in any chosen representation of the variables) with deterministic
changes of some of the degrees of freedom.  Both approaches are interesting, and
advances in either approach for Bayesian CMB analysis could lead to improvements over
the approach presented in this paper.

\end{document}